\documentclass[%
 aip,
 amsmath,amssymb,
 reprint,
]{revtex4-1}

\usepackage{graphicx}
\usepackage{dcolumn}
\usepackage{bm}
\usepackage[utf8]{inputenc}
\usepackage[T1]{fontenc}
\usepackage{mathptmx}
\usepackage{subcaption}

\begin{document}

\preprint{}

\title[Spin-torque Dynamics for Noise Reduction in Vortex-based Sensors]{Spin-torque Dynamics for Noise Reduction in Vortex-based Sensors}

\author{M. Jotta Garcia}
 \email{mafalda.jotta@cnrs-thales.fr}
\affiliation{ 
Unité Mixte de Physique, CNRS, Thales, Université Paris-Saclay, 91767, Palaiseau, France 
}%

\author{J. Moulin}
\affiliation{%
SPEC, CEA, CNRS, Université Paris-Saclay, CEA Saclay 91191 Gif-sur-Yvette Cedex, France}%

\author{S. Wittrock}%
\affiliation{ 
Unité Mixte de Physique, CNRS, Thales, Université Paris-Saclay, 91767, Palaiseau, France 
}%

\author{S. Tsunegi}
\author{K. Yakushiji}
\author{A. Fukushima}
\author{H. Kubota}
\author{S. Yuasa}
 \affiliation{%
National Institute of Advanced Industrial Science and Technology, Research Center for Emerging Computing Technologies, Tsukuba, Ibaraki 305-8568, Japan
}%

\author{U. Ebels}
\affiliation{Univ. Grenoble Alpes, CEA/IRIG, CNRS, GINP, SPINTEC, 38054 Grenoble, France }
\author{M.Pannetier-Lecoeur}
\author{C. Fermon}
\affiliation{%
SPEC, CEA, CNRS, Université Paris-Saclay, CEA Saclay 91191 Gif-sur-Yvette Cedex, France}%

\author{R. Lebrun}
\author{P. Bortolotti}
\affiliation{ 
Unité Mixte de Physique, CNRS, Thales, Université Paris-Saclay, 91767, Palaiseau, France 
}%
\author{A. Solignac}
\affiliation{%
SPEC, CEA, CNRS, Université Paris-Saclay, CEA Saclay 91191 Gif-sur-Yvette Cedex, France}%
\author{V. Cros}
\affiliation{ 
Unité Mixte de Physique, CNRS, Thales, Université Paris-Saclay, 91767, Palaiseau, France 
}%

\date{\today}

\begin{abstract}
Performance of magnetoresistive sensors is today mainly limited by their 1/f low-frequency noise. Here, we study this noise component in vortex-based TMR sensors. We compare the noise level in different magnetization configurations of the device, i.e vortex state or uniform parallel or antiparallel states. We find that the vortex state is at least an order of magnitude noisier than the uniform states. Nevertheless, by activating the spin-transfer induced dynamics of the vortex configuration, we observe a reduction of the 1/f noise, close to the values measured in the AP state, as the vortex core has a lower probability of pinning into defect sites. Additionally, by driving the dynamics of the vortex core by a non-resonant rf field or current we demonstrate that the 1/f noise can be further decreased. The ability to reduce the 1/f low-frequency noise in vortex-based devices by leveraging their spin-transfer dynamics thus enhances their applicability in the magnetic sensors' landscape.
\end{abstract}

\maketitle

Magnetoresistive field sensors have a wide range of uses, such as in biomedical applications \cite{Cardoso2017}, in the automotive industry \cite{Liu2019}, robotics \cite{Alfadhel2016}, and smart city technologies like power-grid monitoring \cite{Gao2020} or, navigation \cite{Yang2015}. Figures of merit like detectivity, sensitivity and spatial resolution are used to evaluate the performance of such sensors \cite{Freitas2016,Leitao2015}. At low-frequencies, the 1/f noise component is dominant and is in fact responsible for limiting the device's detectivity, and consequently its performance \cite{Hardner1993,Mazumdar2007}. Tackling this limitation brings about an active research effort to reduce this noise component \cite{Huang2017,Moulin2019}.

Vortex-based devices, in which the free layer exhibits a vortex magnetization distribution in its equilibrium state, are promising magnetic field sensors due to their large linear detection range \cite{Suess2018} and the fact that they show practically no hysteresis in this range. Besides, these devices are often considered as model systems for the study of magnetization dynamics.
In this study, we focus on the investigation of the 1/f noise in a particular type of magnetic sensor based on a vortex magnetic configuration integrated in a magnetic tunnel junction (MTJ) spin torque nano-oscillator (STNO). STNOs present very good rf characteristics for future radio-frequency (rf) devices and applications \cite{Locatelli2014}, such as rf generation \cite{Dussaux2010,Wittrock2019}, detection \cite{Jenkins2016,Menshawy2017} or neuromorphic computing \cite{Torrejon2017,Romera2018}. While the use of vortex-based STNOs for applications such as these referred here has been largely studied, they are newcomers in the magnetic sensor's landscape. 

Here, we study the 1/f low-frequency noise in vortex-based STNOs, in the first instance, to assess their performance as magnetic field sensors. While there have been studies regarding the noise properties in the low offset frequency regime in the dynamical modes of these devices \cite{Wittrock2019,Wittrock2020} (pertaining to the emitted rf signal), their 1/f low-frequency noise, related to the resistance fluctuations, is largely unstudied. Ultimately, we provide some solutions relying on the STNO's functionalities to decrease the devices' 1/f noise as a means to improve their performance as sensors.

The studied magnetic tunnel junction (MTJ) stack is composed of (Si/SiO$_2$) substrate / buffer layer /synthetic antiferromagnet (SAF)/MgO (1)/ FeB (6)/MgO (1)/capping layers (thickness in nanometers). The pinned SAF layer is a PtMn (15)/CoFe$_{29}$ (2.5)/Ru (0.86)/CoFeB (1.6)/CoFe$_{30}$ (2.5) multilayer. The free layer with a magnetic vortex as the ground state is the FeB layer, with a diameter of $350$~nm. The sample has a tunneling magnetoresistance (TMR) of $85\%$ and an average resistance R$_0=60 $ Ohm. An inductive line sits $300$~nm above the magnetic tunnel junction.

\begin{center}
\begin{figure*}[ht]
\includegraphics[width=0.9\textwidth]{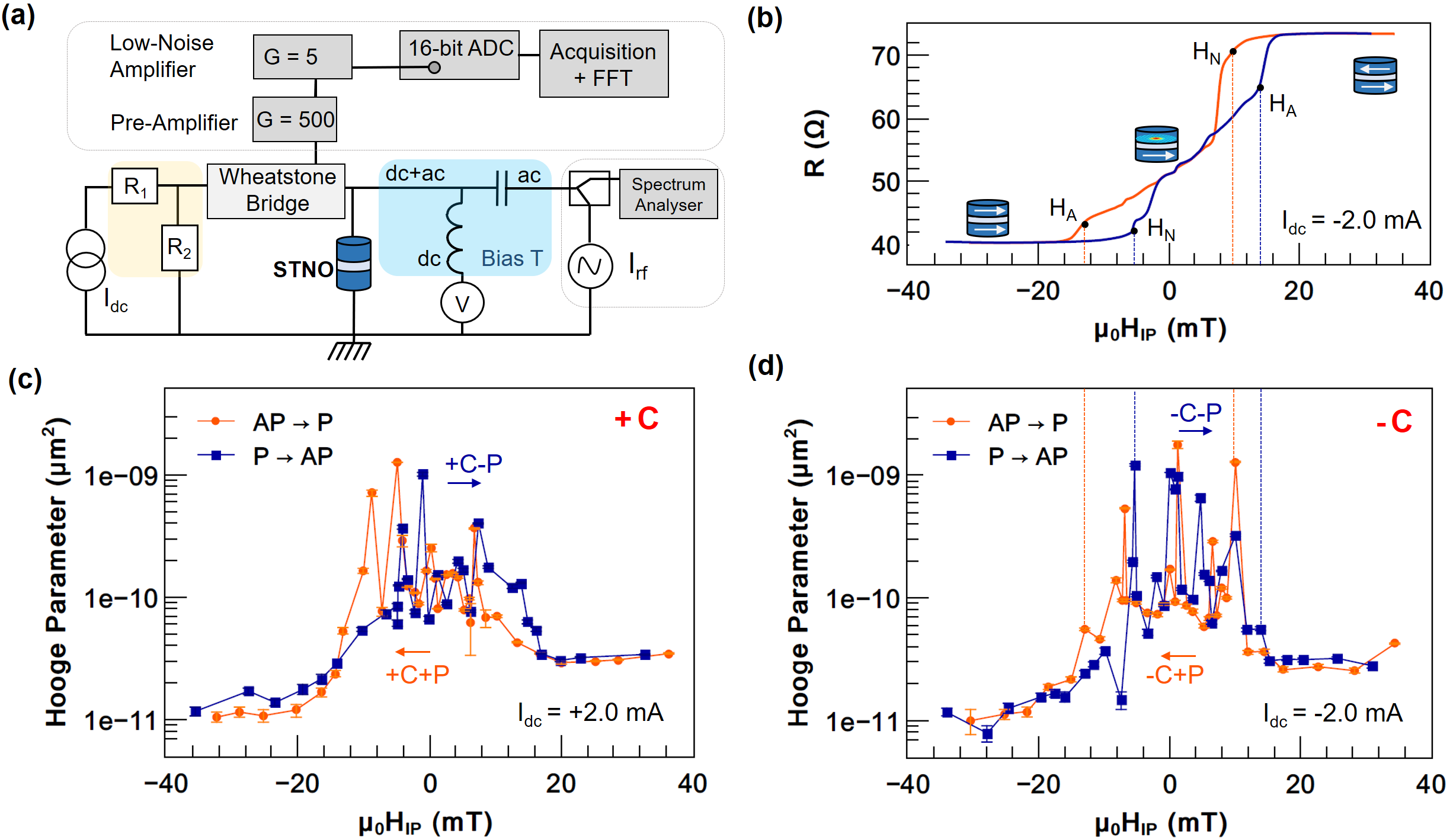}
\caption{(a) Schematic of the measurement setup. (b) Evolution of the STNO device's resistance with the applied field in-plane component, for a $-2.0$~mA bias current, in the negative chirality configuration of the vortex. (c,d) Noise level at low frequency ($1$~Hz-$5$~kHz), represented by the Hooge Parameter, in function of the applied field in-plane component, swept from the anti-parallel (AP) state to the parallel state (P) (in orange) and vice-versa (in blue), for the (c) positive and (d) negative chirality.}
\label{fig:vsH}
\end{figure*}
\end{center}

In Figure \ref{fig:vsH}(a) we describe the measurement circuit that has been designed to allow the simultaneous study of the magnetization dynamics, in the hundreds of MHz range, and the low-frequency noise of the device. The high-frequency component of the circuit consists of a spectrum analyzer and an rf power source. The low-frequency noise measurements were performed by biasing the STNO through a balanced Wheatstone bridge using a dc current source. The output signal is pre-amplified by an INA103 amplifier, followed by a second amplification and filtering chain, reaching a total gain of about $10^3$. The output temporal signal is then acquired by a 16-bit acquisition card. A Fast Fourier Transform (FFT) is performed on the measured signal in order to obtain the noise spectral density, S$_V$. Noise spectral density (NSD) curves typically have a low-frequency $1/f$ component, thermal white noise and lorentzian random telegraph noise (RTN)\cite{Nowak1998,Arakawa2012}. Each noise curve was obtained from averaging over 20 acquisitions and its analysis is done by fitting the different noise components in the range between $1$~Hz and $5000$~Hz. Here, we are interested in the 1/f noise component.

The Hooge parameter $\alpha$, is a commonly used phenomenological parameter~\cite{Hooge1969} used to compare the $1/f$ normalized noise level of different devices with the same RA product. This parameter is extracted from the fitting of the experimental 1/f noise spectral density component, $S_V^{1/f}$, using Eq. \ref{eq:Sv}.

\begin{equation}
\label{eq:Sv}
    S_V^{1/f}=\frac{\alpha V^2}{ A f}
\end{equation}

where $V$ is the average voltage of the device during each measurement, $A$ the device's surface area and $f$ the frequency.\\

The STNO device is placed between the two poles of an electromagnet. We position it at an angle such that the applied field has both in-plane and out-of-plane components, respectively $H_{IP}$ and $H_{OOP}$. 
The vortex magnetization distribution in the studied STNOs is characterized by two parameters, its polarity (P), which is the direction of the vortex’ core magnetization, and its chirality (C), which is the sense of the rotation of the magnetization in the vortex’ body. Vortex-based STNOs present four possible polarity/chirality magnetic configurations. The polarity ($\pm$~P) of the vortex can be set through the application of a large out-of-plane magnetic field, around $\pm 700$~mT. In our experiments, the sign of the out-of-plane field determines the vortex polarity. The vortex chirality ($\pm$~C) can be set through the injection of a large dc current, around $\pm 5$~mA. The direction of the ortho-radial Oersted field generated by the current, which itself depends on the current sign, determines the chirality.

When an in-plane magnetic field is applied, the vortex core is displaced from the disk’s center perpendicularly to the applied field \cite{Guslienko2001}. For a large enough $H_{IP}$ the vortex core reaches the MTJ's edge (also called annihilation field, $H_A$) and the disk’s magnetization becomes uniform, aligning itself with the applied field. In the case where the uniform free layer's magnetization follows the same direction as the fixed reference layer (SAF) - parallel (P) state -, the STNO resistance is the lowest, due to the magnetoresistance effect \cite{Baibich1988,Binasch1989}. Inversely, when the free and fixed layers' magnetization directions oppose each other, the device is in its highest resistance configuration - anti-parallel (AP) state. By decreasing $H_{IP}$, the magnetic vortex is recovered at the nucleation field, $H_N$. As can be seen in Figure \ref{fig:vsH} b), when the device is in the AP (P) state and the applied magnetic field is decreased (increased), there is renucleation of the vortex core in the -C+P (-C-P) configuration.

In order to compare the low-frequency noise in the different magnetic states, the Hooge parameter is determined experimentally, for the positive ($I_{dc}=+2.0$~mA) and negative ($I_{dc}=-2.0$~mA) chirality configurations of the vortex, at different magnetic field values between the P and AP states, passing through the vortex state, and vice-versa (see Figure \ref{fig:vsH}(c,d)). Note that for such $I_{dc}$ values, the STNOs are still in the so-called subcritical regime, meaning that the spin-transfer torques are small enough not to generate a sustained dynamical state of the vortex core. For each magnetic configuration we calculate an average Hooge parameter value. This average value is determined from all the fitted 1/f NSD slopes that were measured in a certain device configuration. The magnetic configuration for each measurement is determined by the resistance of the device (see Figure \ref{fig:vsH}(b)). We find that, for a positive chirality, the vortex state has an average noise level $\alpha_V=2.1\times10^{-10}~\mu$m$^2$ at least one order of magnitude greater than the parallel and anti-parallel states, with Hooge parameters of $\alpha_P=1.4\times10^{-11}~\mu$m$^2$ and $\alpha_{AP}=3.2\times10^{-11}~\mu$m$^2$, respectively (see Figure \ref{fig:vsH}(c)). It is to be noticed that there is a large dispersion of the measured Hooge parameters in the vortex state, of $2.6\times10^{-10}~\mu$m$^2$, which is much larger than the error bars, contrary to what is measured in the saturated cases, where the dispersion is of $0.2\times10^{-11}~\mu$m$^2$ and $0.4\times10^{-11}~\mu$m$^2$, in the P and AP states, respectively. This dispersion is most probably associated with the fact that in the displacement of the vortex core perpendicularly to the applied magnetic field lines \cite{Guslienko2001}, as $H_{IP}$ changes, the vortex core moves between pinning sites and or material grains \cite{Kuepferling2015,Dussaux2010}. When these are present there is an increase in the measured low-frequency noise as it gets pinned. We find that the Hooge parameter in the AP state is threefold that of the P state. This difference is well explained by the electrical $1/f$ noise dependence on the number of open conduction channels in the tunneling barrier, which is higher in the parallel state \cite{Julliere1975}. This is indeed a classical behavior in TMR based sensors \cite{Scola2007}. Hence, in case of a pure electrical origin, it could be expected that in the vortex state the noise would be limited between the parallel and the anti-parallel states' noise levels. Given that we find a noise amplitude much larger than that of the AP one, we elaborate that the magnetic noise component is behind the increase of $\alpha$ in the vortex state, when compared to the saturated states, where the magnetic noise is minimised \cite{Nowak1998}. Interestingly, we find that the measured 1/f noise is independent of the vortex chirality and polarity configuration, given that the Hooge parameter has comparable values in the different configurations (see Figure \ref{fig:vsH} (c,d)).
After having characterized the 1/f noise in the vortex configuration, we propose in the following some strategies to reduce the low-frequency noise of the vortex states close to the values obtained in the uniform states.

\begin{center}
\begin{figure}[h]
\includegraphics[width=0.43\textwidth]{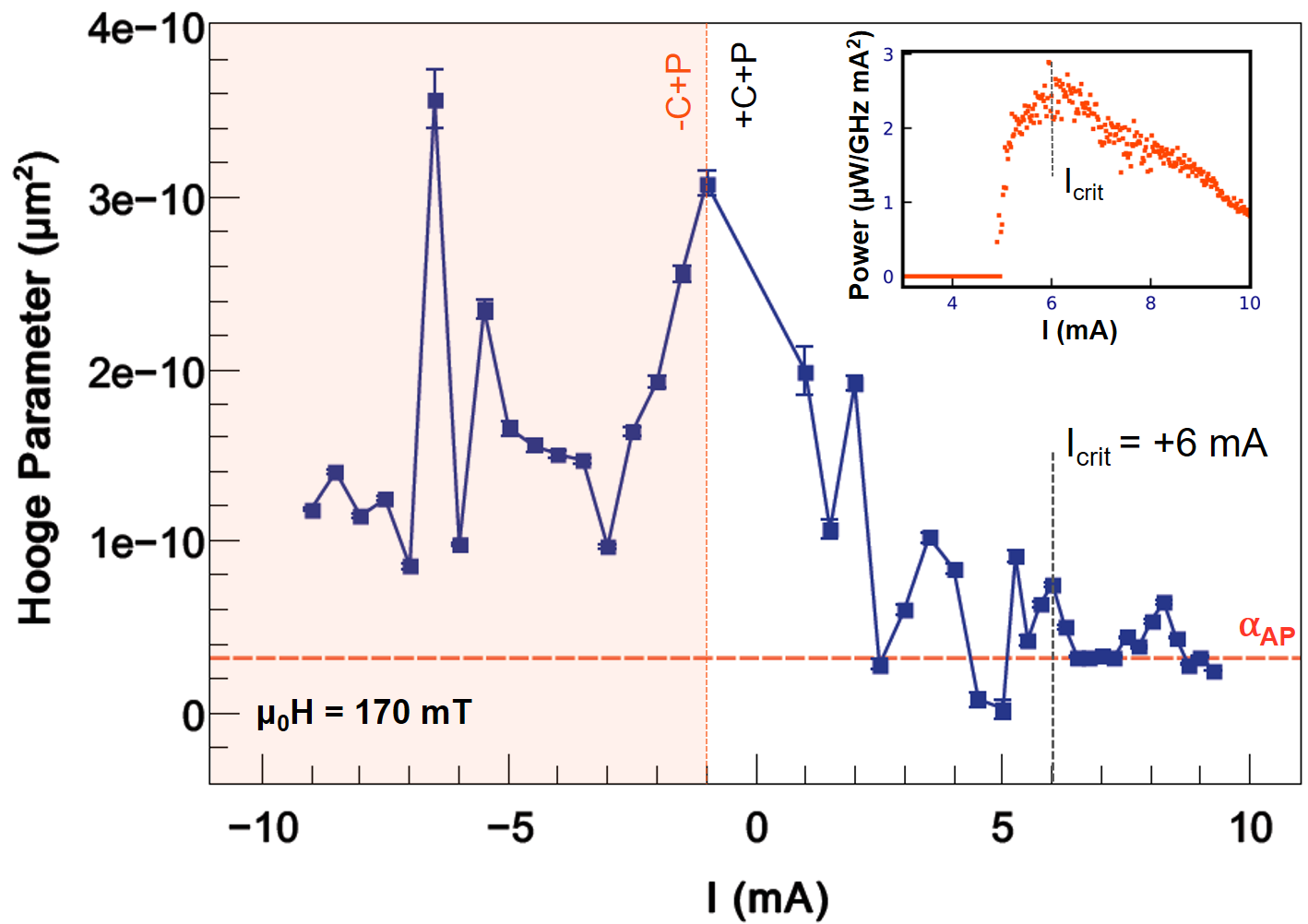}
\caption{Evolution of the Hooge parameter with the applied bias current. The inset shows the oscillation power of the rf emission due to the emerging vortex dynamics. The red dotted line represents the Hooge parameter measured in the AP state.}
\label{fig:vsI}
\end{figure}
\end{center}

A first approach is based on the use of a dc current injected into the STNO device that generates a spin-transfer torque that acts upon the layer's magnetization. For $I_{dc}<0$, the induced spin-transfer torque acts as an extra-damping term and as such, no self-sustained precession of the vortex core occurs. In these measurements, the applied magnetic field is fully out-of-plane, $\mu_0H_{OOP}=170$~mT. As can be observed in Figure \ref{fig:vsI}, we first observe a reduction of the Hooge parameter for $I_{dc}$ between $-1$~mA and $-3~mA$, reaching $\alpha=1.4\times 10^{-10}~\mu$m$^2$, a typical value for the vortex state (see Figure \ref{fig:vsH}).\\

For $I_{dc}>0$, we first see that up to $I_{dc} = 3$~mA the Hooge parameter remains in the range of what is obtained at zero current. For $I_{dc}$ between $3$~mA and $5$~mA, we find that the Hooge parameter gradually decreases. Then, for a large enough current $I_{dc}>I_{crit}$ of $6$~mA, the STNO enters the self-sustained oscillation regime, as can be seen by the increase of the oscillations' power in the inset of Figure \ref{fig:vsI}. While the system's sustained dynamics occur in the radio frequency range, in the case of the studied device around $240$~MHz, we study here how they influence the low-frequency noise of the device. In this regime, we determine a decrease and stabilization of the Hooge parameter value, with the device achieving $\alpha= 3.6\times10^{-11}~\mu$m$^2$. Moreover, we also find a clear decrease of the dispersion of the measured values, reducing to $1.2\times10^{-11}~\mu$m$^2$ (see Figure \ref{fig:vsI}). The precessional movement of the vortex core, in the self-sustained regime, makes it less sensible to material defects of the free layer, therefore decreasing the measured low-frequency noise. This noise reduction is in the magnetic component of the 1/f low-frequency noise.
We find that the vortex magnetization dynamics strongly influence the low-frequency noise of the device. There is a reduction of the 1/f noise of the STNO, while still exhibiting a vortex magnetization distribution at the free layer. The self-sustained oscillations of the vortex don't significantly alter the sensor's large linear detection range, thus keeping its advantage. The measured Hooge parameter in this regime is comparable to that measured in the AP state in the sub-critical regime (see red dotted line in Figure \ref{fig:vsI}).\\

Another approach to improve the 1/f noise amplitude is by relying on the injection of an rf signal into the STNO. In fact, there are two possibilities to generate rf torques acting on the vortex core dynamics, either by using an rf field generated in an rf line close to the device or by using an rf current directly injected into the device. Both these options are tested hereafter.

\begin{center}
\begin{figure}[h]
\includegraphics[width=0.48\textwidth]{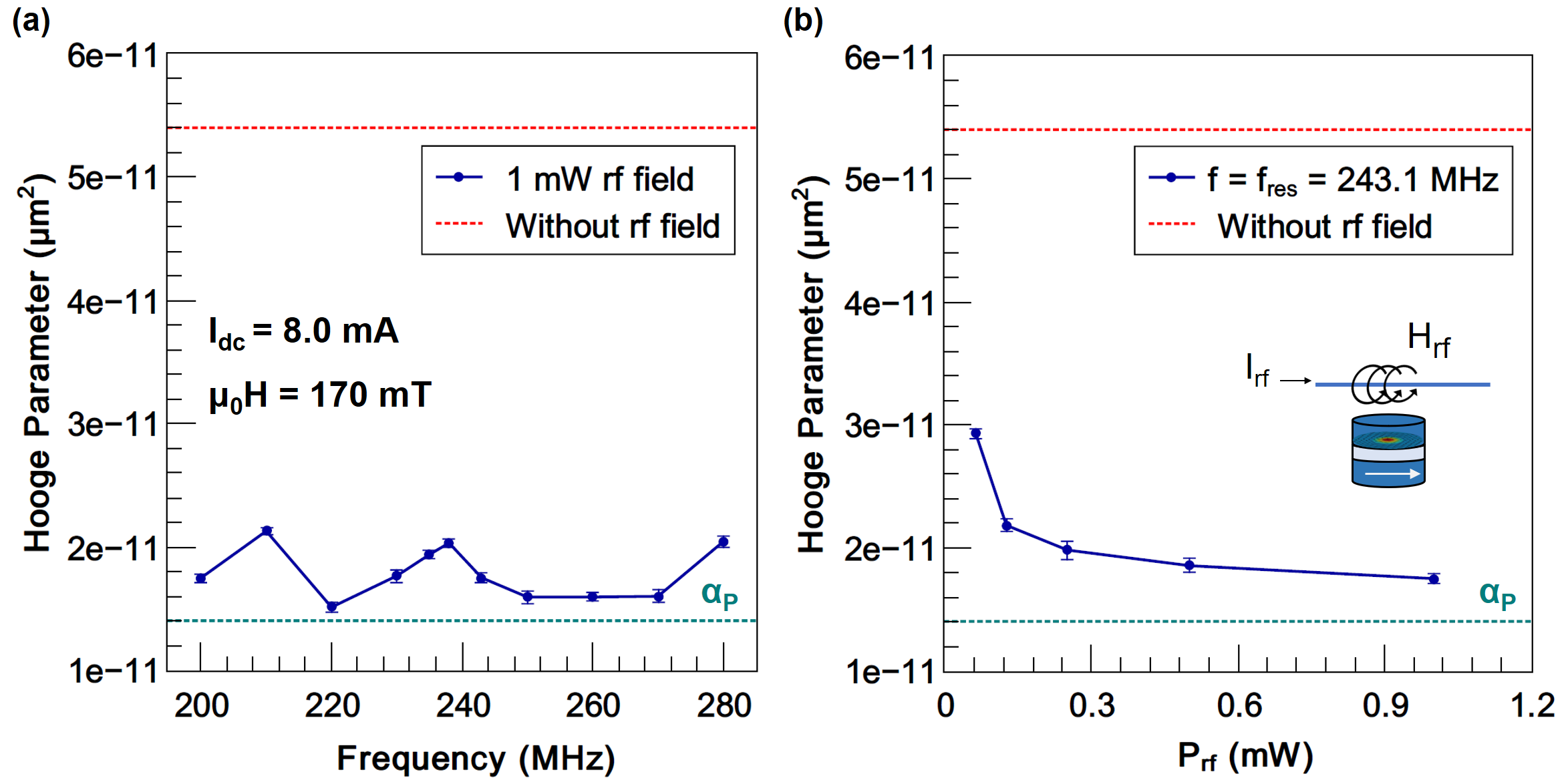}
\caption{(a) Hooge parameter in function of the frequency of the applied oscillating field, with fixed $P_{rf}=1$~mW. The red line indicates the value measured in the absence of this field. (b) Hooge parameter in function of the power amplitude of the field at a fixed frequency, $f=f_{res}=243.1$~MHz. The green dotted lines represent the Hooge parameter measured in the P state.}
\label{fig:B47_rf}
\end{figure}
\end{center}

We study the influence of an alternating magnetic field acting on the vortex magnetization. The injection of an oscillating current into the inductive line generates an oscillating in-plane magnetic field at the free layer. In this case, the device is operating in the self-sustained regime, for $I_{dc}=8.0$~mA and $\mu_0H_{OOP}=170$~mT (see Figure \ref{fig:vsI}). For an injection power of $1$~mW, the rf current amplitude in the inductive line is $I_{rf}=6.9$~mA. For an STNO in the self-sustained operation regime, the measured noise without any rf field is slightly higher than the noise measured in the anti-parallel state, as presented in Figure \ref{fig:vsI} for large positive current. In Figure \ref{fig:B47_rf} (a), we observe that by applying an rf field with a frequency in the range of $200$-$280$~MHz, the noise level at the studied operation conditions (I$_{dc}=8.0$~mA and $\mu_0H_{OOP}=170$~mT) is reduced from $5.4\times10^{-11}~\mu$m$^2$ to a third of this value. The Hooge parameter value measured without an applied rf field in these operation conditions is represented by a red dotted line in Figure \ref{fig:B47_rf}. The average Hooge parameter obtained in this case is $\alpha_{H_{rf}}=1.8\times10^{-11}~\mu$m$^2$. We purposefully chose to sweep a frequency range which includes the STNOs resonance frequency, $243$~MHz. We find that the achieved noise reduction is similar whether the signal is off resonance or in resonance.  

Increasing the intensity of the rf field, the decrease of the 1/f noise level is more pronounced. As the vortex core movement is faster, with less probability of pinning, we find a decrease of the measured Hooge parameter, as shown in Figure~\ref{fig:B47_rf}(b). This noise reduction is limited by the noise level of the parallel state, which is the minimum achievable noise level of the device, represented by a green dotted line in Figure~\ref{fig:B47_rf}. With this strategy, the device's detectivity can be improved - a fundamental factor for magnetic field sensors.\\

A second approach investigated here to drive the dynamics of the vortex system is by directly injecting an rf current, $I_{rf}$, into the STNO, while keeping $I_{dc}<I_{crit}$, so that the STNO remains in the sub-critical (damped) regime. Note that this second series of measurements has been performed on a different STNO from the same wafer but having comparable operation and noise properties. When a $3.2$~nW rf current is injected we measure $\alpha_{I_{rf}}=1.8\times10^{-10}~\mu$m$^2$, while without an rf current we have $\alpha_{V}=3.0\times10^{-10}~\mu$m$^2$. Although there is a reduction of the noise level, the Hooge parameter is still an order of magnitude larger than $\alpha_{P}$, due to the absence of self-sustained oscillations of the vortex core. We observe this decrease for $I_{rf}$ with frequencies close to the nano-oscillators resonance frequency - around $290$~MHz -, but also below it, down to $500$~kHz which is the lower frequency limit of the instruments used in the experimental work. We find that the noise reduction deriving from the rf driven vortex core motion is a non-resonant effect. Compared to the situation where an rf field is applied, much lower rf powers are necessary for the same relative reduction of the noise level, with a few nW being supplied in this case versus slightly below $0.1$~mW in the previous case. This is due to the increased efficiency of the rf current in driving the vortex core motion.

\begin{center}
\begin{figure}[h]
\includegraphics[width=0.43\textwidth]{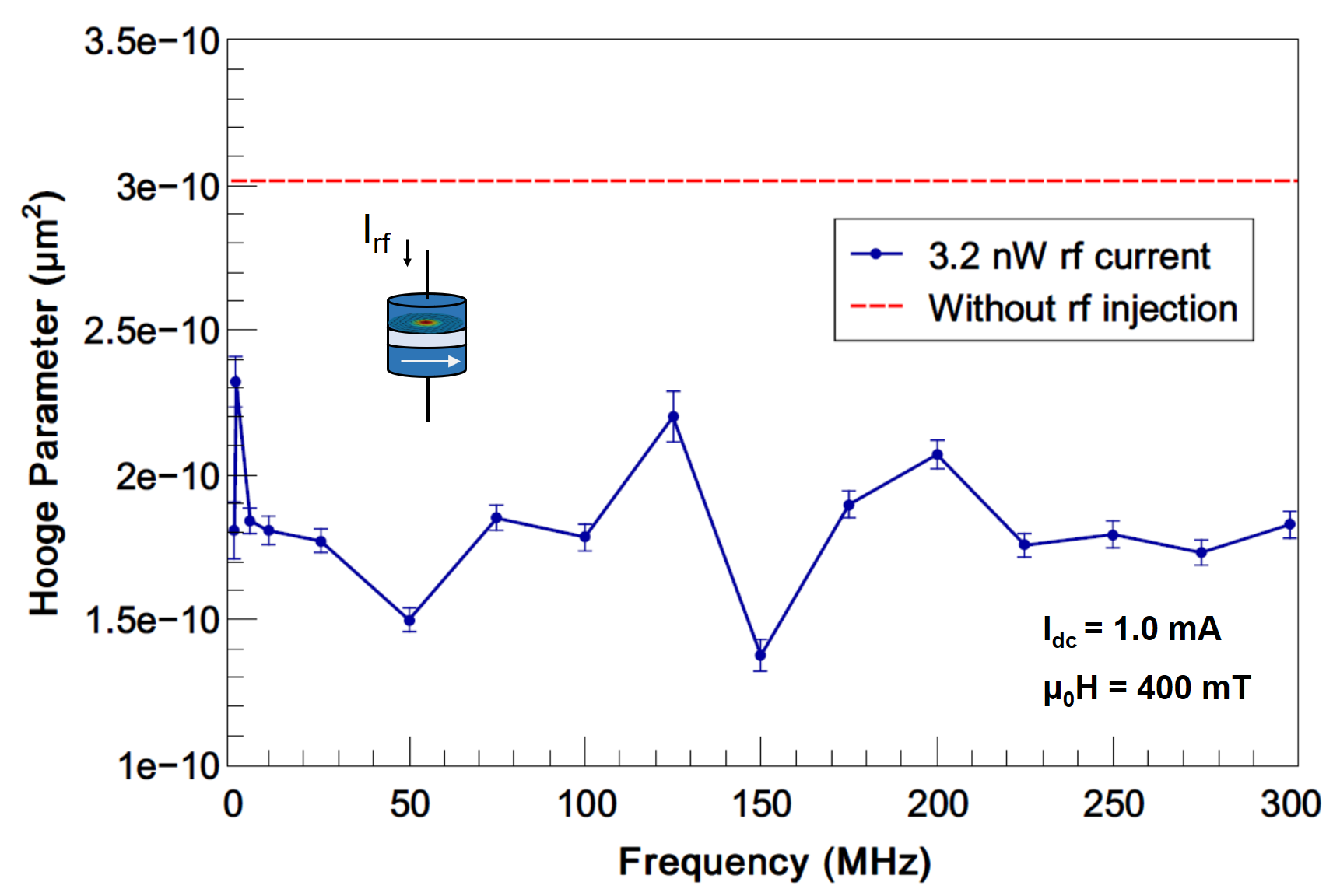}
\caption{Hooge parameter as a function of the injected rf current frequency, for the operating conditions: $\mu_0 H_{IP}=400$~mT, $I_{dc}=1.0$~mA and $P_{rf}=~3.2$nW.}
\label{fig:B08_Irf}
\end{figure}
\end{center} 
In summary, we analyse the 1/f low-frequency noise in vortex-based spin-torque nano-oscillators by determining the Hooge parameter, $\alpha$, in different conditions. Firstly, we find that in the uniform states the $\alpha$ of the studied device is comparable to that of typical state-of-the-art TMR sensors, while in the vortex state it is over one order of magnitude larger. This is due to the increased probability of pinning of the vortex core into defects or inhomogeneities of the free layer. Secondly, we determine that the dynamics of the vortex core strongly influence the noise level of the device. In the self-sustained oscillations' regime the noise decreases to a level close to that of the AP state. 

Furthermore, we present a novel strategy for reducing the 1/f low-frequency magnetic noise, through the application of an in-plane rf field or injection of an rf current. By using this approach while the device is operating in the self-sustained regime, we are capable of further decreasing the measured noise level to values close to the minimum attainable. As such, we can have a vortex-based STNO with relevant noise properties, comparable to those of state-of-the-art TMR field sensors. At the same time, we profit from the specific advantages of vortex-based STNOs for sensing applications: large linear detection range and high spatial resolution. This noise reduction technique based on the spin-torque dynamics of the vortex can have an impact on the sensors' industry, which may profit from the advantages of the vortex configuration.

\begin{acknowledgments}
The work is supported by the French ANR projects “SPINNET” ANR-18-CE24-0012 and "CARAMEL" ANR-18-CE42-0001.
\end{acknowledgments}

\section*{Data availability}
The data that support the findings of this study are available from the corresponding author upon reasonable request.

\bibliography{main}

\providecommand{\noopsort}[1]{}\providecommand{\singleletter}[1]{#1}%
\begin{thebibliography}{29}%
\makeatletter
\providecommand \@ifxundefined [1]{%
 \@ifx{#1\undefined}
}%
\providecommand \@ifnum [1]{%
 \ifnum #1\expandafter \@firstoftwo
 \else \expandafter \@secondoftwo
 \fi
}%
\providecommand \@ifx [1]{%
 \ifx #1\expandafter \@firstoftwo
 \else \expandafter \@secondoftwo
 \fi
}%
\providecommand \natexlab [1]{#1}%
\providecommand \enquote  [1]{``#1''}%
\providecommand \bibnamefont  [1]{#1}%
\providecommand \bibfnamefont [1]{#1}%
\providecommand \citenamefont [1]{#1}%
\providecommand \href@noop [0]{\@secondoftwo}%
\providecommand \href [0]{\begingroup \@sanitize@url \@href}%
\providecommand \@href[1]{\@@startlink{#1}\@@href}%
\providecommand \@@href[1]{\endgroup#1\@@endlink}%
\providecommand \@sanitize@url [0]{\catcode `\\12\catcode `\$12\catcode
  `\&12\catcode `\#12\catcode `\^12\catcode `\_12\catcode `\%12\relax}%
\providecommand \@@startlink[1]{}%
\providecommand \@@endlink[0]{}%
\providecommand \url  [0]{\begingroup\@sanitize@url \@url }%
\providecommand \@url [1]{\endgroup\@href {#1}{\urlprefix }}%
\providecommand \urlprefix  [0]{URL }%
\providecommand \Eprint [0]{\href }%
\providecommand \doibase [0]{http://dx.doi.org/}%
\providecommand \selectlanguage [0]{\@gobble}%
\providecommand \bibinfo  [0]{\@secondoftwo}%
\providecommand \bibfield  [0]{\@secondoftwo}%
\providecommand \translation [1]{[#1]}%
\providecommand \BibitemOpen [0]{}%
\providecommand \bibitemStop [0]{}%
\providecommand \bibitemNoStop [0]{.\EOS\space}%
\providecommand \EOS [0]{\spacefactor3000\relax}%
\providecommand \BibitemShut  [1]{\csname bibitem#1\endcsname}%
\let\auto@bib@innerbib\@empty
\bibitem [{\citenamefont {Cardoso}\ \emph {et~al.}(2017)\citenamefont
  {Cardoso}, \citenamefont {Leitao}, \citenamefont {Dias}, \citenamefont
  {Valadeiro}, \citenamefont {Silva}, \citenamefont {Chícharo}, \citenamefont
  {Silverio}, \citenamefont {Gaspar},\ and\ \citenamefont
  {Freitas}}]{Cardoso2017}%
  \BibitemOpen
  \bibfield  {author} {\bibinfo {author} {\bibfnamefont {S.}~\bibnamefont
  {Cardoso}}, \bibinfo {author} {\bibfnamefont {D.}~\bibnamefont {Leitao}},
  \bibinfo {author} {\bibfnamefont {T.}~\bibnamefont {Dias}}, \bibinfo {author}
  {\bibfnamefont {J.}~\bibnamefont {Valadeiro}}, \bibinfo {author}
  {\bibfnamefont {M.}~\bibnamefont {Silva}}, \bibinfo {author} {\bibfnamefont
  {A.}~\bibnamefont {Chícharo}}, \bibinfo {author} {\bibfnamefont
  {V.}~\bibnamefont {Silverio}}, \bibinfo {author} {\bibfnamefont
  {J.}~\bibnamefont {Gaspar}}, \ and\ \bibinfo {author} {\bibfnamefont
  {P.}~\bibnamefont {Freitas}},\ }\bibfield  {title} {\enquote {\bibinfo
  {title} {Challenges and trends in magnetic sensor integration with
  microfluidics for biomedical applications},}\ }\href@noop {} {\bibfield
  {journal} {\bibinfo  {journal} {J. Phys. D: Appl. Phys.}\ }\textbf {\bibinfo
  {volume} {50}},\ \bibinfo {pages} {213001} (\bibinfo {year}
  {2017})}\BibitemShut {NoStop}%
\bibitem [{\citenamefont {Liu}, \citenamefont {Liu},\ and\ \citenamefont
  {Pong}(2019)}]{Liu2019}%
  \BibitemOpen
  \bibfield  {author} {\bibinfo {author} {\bibfnamefont {X.}~\bibnamefont
  {Liu}}, \bibinfo {author} {\bibfnamefont {C.}~\bibnamefont {Liu}}, \ and\
  \bibinfo {author} {\bibfnamefont {P.~W.}\ \bibnamefont {Pong}},\ }\bibfield
  {title} {\enquote {\bibinfo {title} {{TMR-Sensor-Array-Based
  Misalignment-Tolerant Wireless Charging Technique for Roadway Electric
  Vehicles}},}\ }\href@noop {} {\bibfield  {journal} {\bibinfo  {journal} {IEEE
  Transactions on Magnetics}\ }\textbf {\bibinfo {volume} {7}},\ \bibinfo
  {pages} {1} (\bibinfo {year} {2019})}\BibitemShut {NoStop}%
\bibitem [{\citenamefont {Alfadhel}\ \emph {et~al.}(2016)\citenamefont
  {Alfadhel}, \citenamefont {Khan}, \citenamefont {Cardoso}, \citenamefont
  {Leitao},\ and\ \citenamefont {Kosel}}]{Alfadhel2016}%
  \BibitemOpen
  \bibfield  {author} {\bibinfo {author} {\bibfnamefont {A.}~\bibnamefont
  {Alfadhel}}, \bibinfo {author} {\bibfnamefont {M.~A.}\ \bibnamefont {Khan}},
  \bibinfo {author} {\bibfnamefont {S.}~\bibnamefont {Cardoso}}, \bibinfo
  {author} {\bibfnamefont {D.}~\bibnamefont {Leitao}}, \ and\ \bibinfo {author}
  {\bibfnamefont {J.}~\bibnamefont {Kosel}},\ }\bibfield  {title} {\enquote
  {\bibinfo {title} {A magnetoresistive tactile sensor for harsh environment
  applications},}\ }\href@noop {} {\bibfield  {journal} {\bibinfo  {journal}
  {Sensors (Basel, Switzerland)}\ }\textbf {\bibinfo {volume} {16}},\ \bibinfo
  {pages} {650} (\bibinfo {year} {2016})}\BibitemShut {NoStop}%
\bibitem [{\citenamefont {Gao}\ and\ \citenamefont {Liou}(2020)}]{Gao2020}%
  \BibitemOpen
  \bibfield  {author} {\bibinfo {author} {\bibfnamefont {K.}~\bibnamefont
  {Gao}}\ and\ \bibinfo {author} {\bibfnamefont {S.~H.}\ \bibnamefont {Liou}},\
  }\bibfield  {title} {\enquote {\bibinfo {title} {{Practical challenges of
  magnetic sensors based on magnetic tunnel junctions for power grid
  applications}},}\ }\href@noop {} {\bibfield  {journal} {\bibinfo  {journal}
  {IEEE Magnetics Letters}\ }\textbf {\bibinfo {volume} {11}},\ \bibinfo
  {pages} {1} (\bibinfo {year} {2020})}\BibitemShut {NoStop}%
\bibitem [{\citenamefont {Yang}\ and\ \citenamefont {Lei}(2015)}]{Yang2015}%
  \BibitemOpen
  \bibfield  {author} {\bibinfo {author} {\bibfnamefont {B.}~\bibnamefont
  {Yang}}\ and\ \bibinfo {author} {\bibfnamefont {Y.}~\bibnamefont {Lei}},\
  }\bibfield  {title} {\enquote {\bibinfo {title} {{Vehicle detection and
  classification for low-speed congested traffic with anisotropic
  magnetoresistive sensor}},}\ }\href@noop {} {\bibfield  {journal} {\bibinfo
  {journal} {IEEE Sensors Journal}\ }\textbf {\bibinfo {volume} {15}},\
  \bibinfo {pages} {1132} (\bibinfo {year} {2015})}\BibitemShut {NoStop}%
\bibitem [{\citenamefont {Freitas}, \citenamefont {Ferreira},\ and\
  \citenamefont {Cardoso}(2016)}]{Freitas2016}%
  \BibitemOpen
  \bibfield  {author} {\bibinfo {author} {\bibfnamefont {P.~P.}\ \bibnamefont
  {Freitas}}, \bibinfo {author} {\bibfnamefont {R.}~\bibnamefont {Ferreira}}, \
  and\ \bibinfo {author} {\bibfnamefont {S.}~\bibnamefont {Cardoso}},\
  }\bibfield  {title} {\enquote {\bibinfo {title} {{Spintronic Sensors}},}\
  }\href {\doibase 10.1109/JPROC.2016.2578303} {\bibfield  {journal} {\bibinfo
  {journal} {Proceedings of the IEEE}\ }\textbf {\bibinfo {volume} {104}},\
  \bibinfo {pages} {1894} (\bibinfo {year} {2016})}\BibitemShut {NoStop}%
\bibitem [{\citenamefont {Leitao}\ \emph {et~al.}(2015)\citenamefont {Leitao},
  \citenamefont {Silva}, \citenamefont {Paz}, \citenamefont {Ferreira},
  \citenamefont {Cardoso},\ and\ \citenamefont {Freitas}}]{Leitao2015}%
  \BibitemOpen
  \bibfield  {author} {\bibinfo {author} {\bibfnamefont {D.~C.}\ \bibnamefont
  {Leitao}}, \bibinfo {author} {\bibfnamefont {A.~V.}\ \bibnamefont {Silva}},
  \bibinfo {author} {\bibfnamefont {E.}~\bibnamefont {Paz}}, \bibinfo {author}
  {\bibfnamefont {R.}~\bibnamefont {Ferreira}}, \bibinfo {author}
  {\bibfnamefont {S.}~\bibnamefont {Cardoso}}, \ and\ \bibinfo {author}
  {\bibfnamefont {P.~P.}\ \bibnamefont {Freitas}},\ }\bibfield  {title}
  {\enquote {\bibinfo {title} {{Magnetoresistive nanosensors: Controlling
  magnetism at the nanoscale}},}\ }\href@noop {} {\bibfield  {journal}
  {\bibinfo  {journal} {Nanotechnology}\ }\textbf {\bibinfo {volume} {27}},\
  \bibinfo {pages} {045501} (\bibinfo {year} {2015})}\BibitemShut {NoStop}%
\bibitem [{\citenamefont {Hardner}\ \emph {et~al.}(1993)\citenamefont
  {Hardner}, \citenamefont {Weissman}, \citenamefont {Salamon},\ and\
  \citenamefont {Parkin}}]{Hardner1993}%
  \BibitemOpen
  \bibfield  {author} {\bibinfo {author} {\bibfnamefont {H.~T.}\ \bibnamefont
  {Hardner}}, \bibinfo {author} {\bibfnamefont {M.~B.}\ \bibnamefont
  {Weissman}}, \bibinfo {author} {\bibfnamefont {M.~B.}\ \bibnamefont
  {Salamon}}, \ and\ \bibinfo {author} {\bibfnamefont {S.~S.~P.}\ \bibnamefont
  {Parkin}},\ }\bibfield  {title} {\enquote {\bibinfo {title}
  {Fluctuation-dissipation relation for giant magnetoresistive 1/f noise},}\
  }\href {\doibase 10.1103/PhysRevB.48.16156} {\bibfield  {journal} {\bibinfo
  {journal} {Phys. Rev. B}\ }\textbf {\bibinfo {volume} {48}},\ \bibinfo
  {pages} {16156} (\bibinfo {year} {1993})}\BibitemShut {NoStop}%
\bibitem [{\citenamefont {Mazumdar}\ \emph {et~al.}(2007)\citenamefont
  {Mazumdar}, \citenamefont {Liu}, \citenamefont {Schrag}, \citenamefont
  {Carter}, \citenamefont {Shen},\ and\ \citenamefont {Xiao}}]{Mazumdar2007}%
  \BibitemOpen
  \bibfield  {author} {\bibinfo {author} {\bibfnamefont {D.}~\bibnamefont
  {Mazumdar}}, \bibinfo {author} {\bibfnamefont {X.}~\bibnamefont {Liu}},
  \bibinfo {author} {\bibfnamefont {B.~D.}\ \bibnamefont {Schrag}}, \bibinfo
  {author} {\bibfnamefont {M.}~\bibnamefont {Carter}}, \bibinfo {author}
  {\bibfnamefont {W.}~\bibnamefont {Shen}}, \ and\ \bibinfo {author}
  {\bibfnamefont {G.}~\bibnamefont {Xiao}},\ }\bibfield  {title} {\enquote
  {\bibinfo {title} {Low frequency noise in highly sensitive magnetic tunnel
  junctions with (001) {MgO} tunnel barrier},}\ }\href {\doibase
  10.1063/1.2754352} {\bibfield  {journal} {\bibinfo  {journal} {Applied
  Physics Letters}\ }\textbf {\bibinfo {volume} {91}},\ \bibinfo {pages}
  {033507} (\bibinfo {year} {2007})}\BibitemShut {NoStop}%
\bibitem [{\citenamefont {Huang}\ \emph {et~al.}(2017)\citenamefont {Huang},
  \citenamefont {Yuan}, \citenamefont {Tao}, \citenamefont {Wan}, \citenamefont
  {Guo}, \citenamefont {Zhang}, \citenamefont {Yin}, \citenamefont {Feng},
  \citenamefont {Nakano}, \citenamefont {Naganuma}, \citenamefont {Liu},
  \citenamefont {Yan},\ and\ \citenamefont {Han}}]{Huang2017}%
  \BibitemOpen
  \bibfield  {author} {\bibinfo {author} {\bibfnamefont {L.}~\bibnamefont
  {Huang}}, \bibinfo {author} {\bibfnamefont {Z.~H.}\ \bibnamefont {Yuan}},
  \bibinfo {author} {\bibfnamefont {B.~S.}\ \bibnamefont {Tao}}, \bibinfo
  {author} {\bibfnamefont {C.~H.}\ \bibnamefont {Wan}}, \bibinfo {author}
  {\bibfnamefont {P.}~\bibnamefont {Guo}}, \bibinfo {author} {\bibfnamefont
  {Q.~T.}\ \bibnamefont {Zhang}}, \bibinfo {author} {\bibfnamefont
  {L.}~\bibnamefont {Yin}}, \bibinfo {author} {\bibfnamefont {J.~F.}\
  \bibnamefont {Feng}}, \bibinfo {author} {\bibfnamefont {T.}~\bibnamefont
  {Nakano}}, \bibinfo {author} {\bibfnamefont {H.}~\bibnamefont {Naganuma}},
  \bibinfo {author} {\bibfnamefont {H.~F.}\ \bibnamefont {Liu}}, \bibinfo
  {author} {\bibfnamefont {Y.}~\bibnamefont {Yan}}, \ and\ \bibinfo {author}
  {\bibfnamefont {X.~F.}\ \bibnamefont {Han}},\ }\bibfield  {title} {\enquote
  {\bibinfo {title} {Noise suppression and sensitivity manipulation of magnetic
  tunnel junction sensors with soft magnetic
  {Co$_{70.5}$Fe$_{4.5}$Si$_{15}$B$_{10}$} layer},}\ }\href {\doibase
  10.1063/1.4990478} {\bibfield  {journal} {\bibinfo  {journal} {Journal of
  Applied Physics}\ }\textbf {\bibinfo {volume} {122}},\ \bibinfo {pages}
  {113903} (\bibinfo {year} {2017})}\BibitemShut {NoStop}%
\bibitem [{\citenamefont {Moulin}\ \emph {et~al.}(2019)\citenamefont {Moulin},
  \citenamefont {Doll}, \citenamefont {Paul}, \citenamefont
  {Pannetier-Lecoeur}, \citenamefont {Fermon}, \citenamefont
  {Sergeeva-Chollet},\ and\ \citenamefont {Solignac}}]{Moulin2019}%
  \BibitemOpen
  \bibfield  {author} {\bibinfo {author} {\bibfnamefont {J.}~\bibnamefont
  {Moulin}}, \bibinfo {author} {\bibfnamefont {A.}~\bibnamefont {Doll}},
  \bibinfo {author} {\bibfnamefont {E.}~\bibnamefont {Paul}}, \bibinfo {author}
  {\bibfnamefont {M.}~\bibnamefont {Pannetier-Lecoeur}}, \bibinfo {author}
  {\bibfnamefont {C.}~\bibnamefont {Fermon}}, \bibinfo {author} {\bibfnamefont
  {N.}~\bibnamefont {Sergeeva-Chollet}}, \ and\ \bibinfo {author}
  {\bibfnamefont {A.}~\bibnamefont {Solignac}},\ }\bibfield  {title} {\enquote
  {\bibinfo {title} {Optimizing magnetoresistive sensor signal-to-noise via
  pinning field tuning},}\ }\href@noop {} {\bibfield  {journal} {\bibinfo
  {journal} {Applied Physics Letters}\ }\textbf {\bibinfo {volume} {115}},\
  \bibinfo {pages} {122406} (\bibinfo {year} {2019})}\BibitemShut {NoStop}%
\bibitem [{\citenamefont {Suess}\ \emph {et~al.}(2018)\citenamefont {Suess},
  \citenamefont {Bachleitner-Hofmann}, \citenamefont {Satz}, \citenamefont
  {Weitensfelder}, \citenamefont {Vogler}, \citenamefont {Bruckner},
  \citenamefont {Abert}, \citenamefont {Pr{\"{u}}gl}, \citenamefont {Zimmer},
  \citenamefont {Huber}, \citenamefont {Luber}, \citenamefont {Raberg},
  \citenamefont {Schrefl},\ and\ \citenamefont {Br{\"{u}}ckl}}]{Suess2018}%
  \BibitemOpen
  \bibfield  {author} {\bibinfo {author} {\bibfnamefont {D.}~\bibnamefont
  {Suess}}, \bibinfo {author} {\bibfnamefont {A.}~\bibnamefont
  {Bachleitner-Hofmann}}, \bibinfo {author} {\bibfnamefont {A.}~\bibnamefont
  {Satz}}, \bibinfo {author} {\bibfnamefont {H.}~\bibnamefont {Weitensfelder}},
  \bibinfo {author} {\bibfnamefont {C.}~\bibnamefont {Vogler}}, \bibinfo
  {author} {\bibfnamefont {F.}~\bibnamefont {Bruckner}}, \bibinfo {author}
  {\bibfnamefont {C.}~\bibnamefont {Abert}}, \bibinfo {author} {\bibfnamefont
  {K.}~\bibnamefont {Pr{\"{u}}gl}}, \bibinfo {author} {\bibfnamefont
  {J.}~\bibnamefont {Zimmer}}, \bibinfo {author} {\bibfnamefont
  {C.}~\bibnamefont {Huber}}, \bibinfo {author} {\bibfnamefont
  {S.}~\bibnamefont {Luber}}, \bibinfo {author} {\bibfnamefont
  {W.}~\bibnamefont {Raberg}}, \bibinfo {author} {\bibfnamefont
  {T.}~\bibnamefont {Schrefl}}, \ and\ \bibinfo {author} {\bibfnamefont
  {H.}~\bibnamefont {Br{\"{u}}ckl}},\ }\bibfield  {title} {\enquote {\bibinfo
  {title} {{Topologically protected vortex structures for low-noise magnetic
  sensors with high linear range}},}\ }\href@noop {} {\bibfield  {journal}
  {\bibinfo  {journal} {Nature Electronics}\ }\textbf {\bibinfo {volume} {1}},\
  \bibinfo {pages} {362} (\bibinfo {year} {2018})}\BibitemShut {NoStop}%
\bibitem [{\citenamefont {Locatelli}, \citenamefont {Cros},\ and\ \citenamefont
  {Grollier}(2014)}]{Locatelli2014}%
  \BibitemOpen
  \bibfield  {author} {\bibinfo {author} {\bibfnamefont {N.}~\bibnamefont
  {Locatelli}}, \bibinfo {author} {\bibfnamefont {V.}~\bibnamefont {Cros}}, \
  and\ \bibinfo {author} {\bibfnamefont {J.}~\bibnamefont {Grollier}},\
  }\bibfield  {title} {\enquote {\bibinfo {title} {{Spin-torque building
  blocks}},}\ }\href@noop {} {\bibfield  {journal} {\bibinfo  {journal} {Nature
  Materials}\ }\textbf {\bibinfo {volume} {13}},\ \bibinfo {pages} {11}
  (\bibinfo {year} {2014})}\BibitemShut {NoStop}%
\bibitem [{\citenamefont {Dussaux}\ \emph {et~al.}(2010)\citenamefont
  {Dussaux}, \citenamefont {Georges}, \citenamefont {Grollier}, \citenamefont
  {Cros}, \citenamefont {Khvalkovskiy}, \citenamefont {Fukushima},
  \citenamefont {Konoto}, \citenamefont {Kubota}, \citenamefont {Yakushiji},
  \citenamefont {Yuasa}, \citenamefont {Zvezdin}, \citenamefont {Ando},\ and\
  \citenamefont {Fert}}]{Dussaux2010}%
  \BibitemOpen
  \bibfield  {author} {\bibinfo {author} {\bibfnamefont {A.}~\bibnamefont
  {Dussaux}}, \bibinfo {author} {\bibfnamefont {B.}~\bibnamefont {Georges}},
  \bibinfo {author} {\bibfnamefont {J.}~\bibnamefont {Grollier}}, \bibinfo
  {author} {\bibfnamefont {V.}~\bibnamefont {Cros}}, \bibinfo {author}
  {\bibfnamefont {A.~V.}\ \bibnamefont {Khvalkovskiy}}, \bibinfo {author}
  {\bibfnamefont {A.}~\bibnamefont {Fukushima}}, \bibinfo {author}
  {\bibfnamefont {M.}~\bibnamefont {Konoto}}, \bibinfo {author} {\bibfnamefont
  {H.}~\bibnamefont {Kubota}}, \bibinfo {author} {\bibfnamefont
  {K.}~\bibnamefont {Yakushiji}}, \bibinfo {author} {\bibfnamefont
  {S.}~\bibnamefont {Yuasa}}, \bibinfo {author} {\bibfnamefont {K.~A.}\
  \bibnamefont {Zvezdin}}, \bibinfo {author} {\bibfnamefont {K.}~\bibnamefont
  {Ando}}, \ and\ \bibinfo {author} {\bibfnamefont {A.}~\bibnamefont {Fert}},\
  }\bibfield  {title} {\enquote {\bibinfo {title} {{Large microwave generation
  from current-driven magnetic vortex oscillators in magnetic tunnel
  junctions}},}\ }\href@noop {} {\bibfield  {journal} {\bibinfo  {journal}
  {Nature Communications}\ }\textbf {\bibinfo {volume} {1}},\ \bibinfo {pages}
  {8} (\bibinfo {year} {2010})}\BibitemShut {NoStop}%
\bibitem [{\citenamefont {Wittrock}\ \emph {et~al.}(2019)\citenamefont
  {Wittrock}, \citenamefont {Tsunegi}, \citenamefont {Yakushiji}, \citenamefont
  {Fukushima}, \citenamefont {Kubota}, \citenamefont {Bortolotti},
  \citenamefont {Ebels}, \citenamefont {Yuasa}, \citenamefont {Cibiel},
  \citenamefont {Galliou}, \citenamefont {Rubiola},\ and\ \citenamefont
  {Cros}}]{Wittrock2019}%
  \BibitemOpen
  \bibfield  {author} {\bibinfo {author} {\bibfnamefont {S.}~\bibnamefont
  {Wittrock}}, \bibinfo {author} {\bibfnamefont {S.}~\bibnamefont {Tsunegi}},
  \bibinfo {author} {\bibfnamefont {K.}~\bibnamefont {Yakushiji}}, \bibinfo
  {author} {\bibfnamefont {A.}~\bibnamefont {Fukushima}}, \bibinfo {author}
  {\bibfnamefont {H.}~\bibnamefont {Kubota}}, \bibinfo {author} {\bibfnamefont
  {P.}~\bibnamefont {Bortolotti}}, \bibinfo {author} {\bibfnamefont
  {U.}~\bibnamefont {Ebels}}, \bibinfo {author} {\bibfnamefont
  {S.}~\bibnamefont {Yuasa}}, \bibinfo {author} {\bibfnamefont
  {G.}~\bibnamefont {Cibiel}}, \bibinfo {author} {\bibfnamefont
  {S.}~\bibnamefont {Galliou}}, \bibinfo {author} {\bibfnamefont
  {E.}~\bibnamefont {Rubiola}}, \ and\ \bibinfo {author} {\bibfnamefont
  {V.}~\bibnamefont {Cros}},\ }\bibfield  {title} {\enquote {\bibinfo {title}
  {{Low offset frequency 1/f flicker noise in spin-torque vortex
  oscillators}},}\ }\href@noop {} {\bibfield  {journal} {\bibinfo  {journal}
  {Physical Review B}\ }\textbf {\bibinfo {volume} {99}},\ \bibinfo {pages}
  {235135} (\bibinfo {year} {2019})}\BibitemShut {NoStop}%
\bibitem [{\citenamefont {Jenkins}\ \emph {et~al.}(2016)\citenamefont
  {Jenkins}, \citenamefont {Lebrun}, \citenamefont {Grimaldi}, \citenamefont
  {Tsunegi}, \citenamefont {Bortolotti}, \citenamefont {Kubota}, \citenamefont
  {Yakushiji}, \citenamefont {Fukushima}, \citenamefont {{De Loubens}},
  \citenamefont {Klein}, \citenamefont {Yuasa},\ and\ \citenamefont
  {Cros}}]{Jenkins2016}%
  \BibitemOpen
  \bibfield  {author} {\bibinfo {author} {\bibfnamefont {A.~S.}\ \bibnamefont
  {Jenkins}}, \bibinfo {author} {\bibfnamefont {R.}~\bibnamefont {Lebrun}},
  \bibinfo {author} {\bibfnamefont {E.}~\bibnamefont {Grimaldi}}, \bibinfo
  {author} {\bibfnamefont {S.}~\bibnamefont {Tsunegi}}, \bibinfo {author}
  {\bibfnamefont {P.}~\bibnamefont {Bortolotti}}, \bibinfo {author}
  {\bibfnamefont {H.}~\bibnamefont {Kubota}}, \bibinfo {author} {\bibfnamefont
  {K.}~\bibnamefont {Yakushiji}}, \bibinfo {author} {\bibfnamefont
  {A.}~\bibnamefont {Fukushima}}, \bibinfo {author} {\bibfnamefont
  {G.}~\bibnamefont {{De Loubens}}}, \bibinfo {author} {\bibfnamefont
  {O.}~\bibnamefont {Klein}}, \bibinfo {author} {\bibfnamefont
  {S.}~\bibnamefont {Yuasa}}, \ and\ \bibinfo {author} {\bibfnamefont
  {V.}~\bibnamefont {Cros}},\ }\bibfield  {title} {\enquote {\bibinfo {title}
  {{Spin-torque resonant expulsion of the vortex core for an efficient
  radiofrequency detection scheme}},}\ }\href@noop {} {\bibfield  {journal}
  {\bibinfo  {journal} {Nature Nanotechnology}\ }\textbf {\bibinfo {volume}
  {11}},\ \bibinfo {pages} {360} (\bibinfo {year} {2016})}\BibitemShut
  {NoStop}%
\bibitem [{\citenamefont {Menshawy}\ \emph {et~al.}(2017)\citenamefont
  {Menshawy}, \citenamefont {Jenkins}, \citenamefont {Merazzo}, \citenamefont
  {Vila}, \citenamefont {Ferreira}, \citenamefont {Cyrille}, \citenamefont
  {Ebels}, \citenamefont {Bortolotti}, \citenamefont {Kermorvant},\ and\
  \citenamefont {Cros}}]{Menshawy2017}%
  \BibitemOpen
  \bibfield  {author} {\bibinfo {author} {\bibfnamefont {S.}~\bibnamefont
  {Menshawy}}, \bibinfo {author} {\bibfnamefont {A.~S.}\ \bibnamefont
  {Jenkins}}, \bibinfo {author} {\bibfnamefont {K.~J.}\ \bibnamefont
  {Merazzo}}, \bibinfo {author} {\bibfnamefont {L.}~\bibnamefont {Vila}},
  \bibinfo {author} {\bibfnamefont {R.}~\bibnamefont {Ferreira}}, \bibinfo
  {author} {\bibfnamefont {M.~C.}\ \bibnamefont {Cyrille}}, \bibinfo {author}
  {\bibfnamefont {U.}~\bibnamefont {Ebels}}, \bibinfo {author} {\bibfnamefont
  {P.}~\bibnamefont {Bortolotti}}, \bibinfo {author} {\bibfnamefont
  {J.}~\bibnamefont {Kermorvant}}, \ and\ \bibinfo {author} {\bibfnamefont
  {V.}~\bibnamefont {Cros}},\ }\bibfield  {title} {\enquote {\bibinfo {title}
  {{Spin transfer driven resonant expulsion of a magnetic vortex core for
  efficient rf detector}},}\ }\href@noop {} {\bibfield  {journal} {\bibinfo
  {journal} {AIP Advances}\ }\textbf {\bibinfo {volume} {7}},\ \bibinfo {pages}
  {056608} (\bibinfo {year} {2017})}\BibitemShut {NoStop}%
\bibitem [{\citenamefont {Torrejon}\ \emph {et~al.}(2017)\citenamefont
  {Torrejon}, \citenamefont {Riou}, \citenamefont {Araujo}, \citenamefont
  {Tsunegi}, \citenamefont {Khalsa}, \citenamefont {Querlioz}, \citenamefont
  {Bortolotti}, \citenamefont {Cros}, \citenamefont {Yakushiji}, \citenamefont
  {Fukushima}, \citenamefont {Kubota}, \citenamefont {Yuasa}, \citenamefont
  {Stiles},\ and\ \citenamefont {Grollier}}]{Torrejon2017}%
  \BibitemOpen
  \bibfield  {author} {\bibinfo {author} {\bibfnamefont {J.}~\bibnamefont
  {Torrejon}}, \bibinfo {author} {\bibfnamefont {M.}~\bibnamefont {Riou}},
  \bibinfo {author} {\bibfnamefont {F.~A.}\ \bibnamefont {Araujo}}, \bibinfo
  {author} {\bibfnamefont {S.}~\bibnamefont {Tsunegi}}, \bibinfo {author}
  {\bibfnamefont {G.}~\bibnamefont {Khalsa}}, \bibinfo {author} {\bibfnamefont
  {D.}~\bibnamefont {Querlioz}}, \bibinfo {author} {\bibfnamefont
  {P.}~\bibnamefont {Bortolotti}}, \bibinfo {author} {\bibfnamefont
  {V.}~\bibnamefont {Cros}}, \bibinfo {author} {\bibfnamefont {K.}~\bibnamefont
  {Yakushiji}}, \bibinfo {author} {\bibfnamefont {A.}~\bibnamefont
  {Fukushima}}, \bibinfo {author} {\bibfnamefont {H.}~\bibnamefont {Kubota}},
  \bibinfo {author} {\bibfnamefont {S.}~\bibnamefont {Yuasa}}, \bibinfo
  {author} {\bibfnamefont {M.~D.}\ \bibnamefont {Stiles}}, \ and\ \bibinfo
  {author} {\bibfnamefont {J.}~\bibnamefont {Grollier}},\ }\bibfield  {title}
  {\enquote {\bibinfo {title} {{Neuromorphic computing with nanoscale
  spintronic oscillators}},}\ }\href@noop {} {\bibfield  {journal} {\bibinfo
  {journal} {Nature}\ }\textbf {\bibinfo {volume} {547}},\ \bibinfo {pages}
  {428} (\bibinfo {year} {2017})}\BibitemShut {NoStop}%
\bibitem [{\citenamefont {Romera}\ \emph {et~al.}(2018)\citenamefont {Romera},
  \citenamefont {Talatchian}, \citenamefont {Tsunegi}, \citenamefont {{Abreu
  Araujo}}, \citenamefont {Cros}, \citenamefont {Bortolotti}, \citenamefont
  {Trastoy}, \citenamefont {Yakushiji}, \citenamefont {Fukushima},
  \citenamefont {Kubota}, \citenamefont {Yuasa}, \citenamefont {Ernoult},
  \citenamefont {Vodenicarevic}, \citenamefont {Hirtzlin}, \citenamefont
  {Locatelli}, \citenamefont {Querlioz},\ and\ \citenamefont
  {Grollier}}]{Romera2018}%
  \BibitemOpen
  \bibfield  {author} {\bibinfo {author} {\bibfnamefont {M.}~\bibnamefont
  {Romera}}, \bibinfo {author} {\bibfnamefont {P.}~\bibnamefont {Talatchian}},
  \bibinfo {author} {\bibfnamefont {S.}~\bibnamefont {Tsunegi}}, \bibinfo
  {author} {\bibfnamefont {F.}~\bibnamefont {{Abreu Araujo}}}, \bibinfo
  {author} {\bibfnamefont {V.}~\bibnamefont {Cros}}, \bibinfo {author}
  {\bibfnamefont {P.}~\bibnamefont {Bortolotti}}, \bibinfo {author}
  {\bibfnamefont {J.}~\bibnamefont {Trastoy}}, \bibinfo {author} {\bibfnamefont
  {K.}~\bibnamefont {Yakushiji}}, \bibinfo {author} {\bibfnamefont
  {A.}~\bibnamefont {Fukushima}}, \bibinfo {author} {\bibfnamefont
  {H.}~\bibnamefont {Kubota}}, \bibinfo {author} {\bibfnamefont
  {S.}~\bibnamefont {Yuasa}}, \bibinfo {author} {\bibfnamefont
  {M.}~\bibnamefont {Ernoult}}, \bibinfo {author} {\bibfnamefont
  {D.}~\bibnamefont {Vodenicarevic}}, \bibinfo {author} {\bibfnamefont
  {T.}~\bibnamefont {Hirtzlin}}, \bibinfo {author} {\bibfnamefont
  {N.}~\bibnamefont {Locatelli}}, \bibinfo {author} {\bibfnamefont
  {D.}~\bibnamefont {Querlioz}}, \ and\ \bibinfo {author} {\bibfnamefont
  {J.}~\bibnamefont {Grollier}},\ }\bibfield  {title} {\enquote {\bibinfo
  {title} {{Vowel recognition with four coupled spin-torque
  nano-oscillators}},}\ }\href@noop {} {\bibfield  {journal} {\bibinfo
  {journal} {Nature}\ }\textbf {\bibinfo {volume} {563}},\ \bibinfo {pages}
  {230} (\bibinfo {year} {2018})}\BibitemShut {NoStop}%
\bibitem [{\citenamefont {Wittrock}\ \emph {et~al.}(2020)\citenamefont
  {Wittrock}, \citenamefont {Talatchian}, \citenamefont {Tsunegi},
  \citenamefont {Cr{\'{e}}t{\'{e}}}, \citenamefont {Yakushiji}, \citenamefont
  {Bortolotti}, \citenamefont {Ebels}, \citenamefont {Fukushima}, \citenamefont
  {Kubota}, \citenamefont {Yuasa}, \citenamefont {Grollier}, \citenamefont
  {Cibiel}, \citenamefont {Galliou}, \citenamefont {Rubiola},\ and\
  \citenamefont {Cros}}]{Wittrock2020}%
  \BibitemOpen
  \bibfield  {author} {\bibinfo {author} {\bibfnamefont {S.}~\bibnamefont
  {Wittrock}}, \bibinfo {author} {\bibfnamefont {P.}~\bibnamefont
  {Talatchian}}, \bibinfo {author} {\bibfnamefont {S.}~\bibnamefont {Tsunegi}},
  \bibinfo {author} {\bibfnamefont {D.}~\bibnamefont {Cr{\'{e}}t{\'{e}}}},
  \bibinfo {author} {\bibfnamefont {K.}~\bibnamefont {Yakushiji}}, \bibinfo
  {author} {\bibfnamefont {P.}~\bibnamefont {Bortolotti}}, \bibinfo {author}
  {\bibfnamefont {U.}~\bibnamefont {Ebels}}, \bibinfo {author} {\bibfnamefont
  {A.}~\bibnamefont {Fukushima}}, \bibinfo {author} {\bibfnamefont
  {H.}~\bibnamefont {Kubota}}, \bibinfo {author} {\bibfnamefont
  {S.}~\bibnamefont {Yuasa}}, \bibinfo {author} {\bibfnamefont
  {J.}~\bibnamefont {Grollier}}, \bibinfo {author} {\bibfnamefont
  {G.}~\bibnamefont {Cibiel}}, \bibinfo {author} {\bibfnamefont
  {S.}~\bibnamefont {Galliou}}, \bibinfo {author} {\bibfnamefont
  {E.}~\bibnamefont {Rubiola}}, \ and\ \bibinfo {author} {\bibfnamefont
  {V.}~\bibnamefont {Cros}},\ }\bibfield  {title} {\enquote {\bibinfo {title}
  {{Influence of flicker noise and nonlinearity on the frequency spectrum of
  spin torque nano-oscillators}},}\ }\href@noop {} {\bibfield  {journal}
  {\bibinfo  {journal} {Scientific Reports}\ }\textbf {\bibinfo {volume}
  {10}},\ \bibinfo {pages} {1} (\bibinfo {year} {2020})}\BibitemShut {NoStop}%
\bibitem [{\citenamefont {Nowak}\ \emph {et~al.}(1998)\citenamefont {Nowak},
  \citenamefont {Merithew}, \citenamefont {Weissman}, \citenamefont {Bloom},\
  and\ \citenamefont {Parkin}}]{Nowak1998}%
  \BibitemOpen
  \bibfield  {author} {\bibinfo {author} {\bibfnamefont {E.~R.}\ \bibnamefont
  {Nowak}}, \bibinfo {author} {\bibfnamefont {R.~D.}\ \bibnamefont {Merithew}},
  \bibinfo {author} {\bibfnamefont {M.~B.}\ \bibnamefont {Weissman}}, \bibinfo
  {author} {\bibfnamefont {I.}~\bibnamefont {Bloom}}, \ and\ \bibinfo {author}
  {\bibfnamefont {S.~S.}\ \bibnamefont {Parkin}},\ }\bibfield  {title}
  {\enquote {\bibinfo {title} {{Noise properties of ferromagnetic tunnel
  junctions}},}\ }\href@noop {} {\bibfield  {journal} {\bibinfo  {journal}
  {Journal of Applied Physics}\ }\textbf {\bibinfo {volume} {84}},\ \bibinfo
  {pages} {6195} (\bibinfo {year} {1998})}\BibitemShut {NoStop}%
\bibitem [{\citenamefont {Arakawa}\ \emph {et~al.}(2012)\citenamefont
  {Arakawa}, \citenamefont {Tanaka}, \citenamefont {Chida}, \citenamefont
  {Matsuo}, \citenamefont {Nishihara}, \citenamefont {Chiba}, \citenamefont
  {Kobayashi}, \citenamefont {Ono}, \citenamefont {Fukushima},\ and\
  \citenamefont {Yuasa}}]{Arakawa2012}%
  \BibitemOpen
  \bibfield  {author} {\bibinfo {author} {\bibfnamefont {T.}~\bibnamefont
  {Arakawa}}, \bibinfo {author} {\bibfnamefont {T.}~\bibnamefont {Tanaka}},
  \bibinfo {author} {\bibfnamefont {K.}~\bibnamefont {Chida}}, \bibinfo
  {author} {\bibfnamefont {S.}~\bibnamefont {Matsuo}}, \bibinfo {author}
  {\bibfnamefont {Y.}~\bibnamefont {Nishihara}}, \bibinfo {author}
  {\bibfnamefont {D.}~\bibnamefont {Chiba}}, \bibinfo {author} {\bibfnamefont
  {K.}~\bibnamefont {Kobayashi}}, \bibinfo {author} {\bibfnamefont
  {T.}~\bibnamefont {Ono}}, \bibinfo {author} {\bibfnamefont {A.}~\bibnamefont
  {Fukushima}}, \ and\ \bibinfo {author} {\bibfnamefont {S.}~\bibnamefont
  {Yuasa}},\ }\bibfield  {title} {\enquote {\bibinfo {title} {{Low-frequency
  and shot noises in {CoFeB/MgO/CoFeB} magnetic tunneling junctions}},}\
  }\href@noop {} {\bibfield  {journal} {\bibinfo  {journal} {Physical Review B
  - Condensed Matter and Materials Physics}\ }\textbf {\bibinfo {volume}
  {86}},\ \bibinfo {pages} {224423} (\bibinfo {year} {2012})}\BibitemShut
  {NoStop}%
\bibitem [{\citenamefont {Hooge}\ and\ \citenamefont
  {Hoppenbrouwers}(1969)}]{Hooge1969}%
  \BibitemOpen
  \bibfield  {author} {\bibinfo {author} {\bibfnamefont {F.}~\bibnamefont
  {Hooge}}\ and\ \bibinfo {author} {\bibfnamefont {A.}~\bibnamefont
  {Hoppenbrouwers}},\ }\bibfield  {title} {\enquote {\bibinfo {title} {1/f
  noise in continuous thin gold films},}\ }\href {\doibase
  10.1016/0031-8914(69)90266-3} {\bibfield  {journal} {\bibinfo  {journal}
  {Physica}\ }\textbf {\bibinfo {volume} {45}},\ \bibinfo {pages} {386}
  (\bibinfo {year} {1969})}\BibitemShut {NoStop}%
\bibitem [{\citenamefont {Guslienko}\ \emph {et~al.}(2001)\citenamefont
  {Guslienko}, \citenamefont {Novosad}, \citenamefont {Otani}, \citenamefont
  {Shima},\ and\ \citenamefont {Fukamichi}}]{Guslienko2001}%
  \BibitemOpen
  \bibfield  {author} {\bibinfo {author} {\bibfnamefont {K.~Y.}\ \bibnamefont
  {Guslienko}}, \bibinfo {author} {\bibfnamefont {V.}~\bibnamefont {Novosad}},
  \bibinfo {author} {\bibfnamefont {Y.}~\bibnamefont {Otani}}, \bibinfo
  {author} {\bibfnamefont {H.}~\bibnamefont {Shima}}, \ and\ \bibinfo {author}
  {\bibfnamefont {K.}~\bibnamefont {Fukamichi}},\ }\bibfield  {title} {\enquote
  {\bibinfo {title} {{Field evolution of magnetic vortex state in ferromagnetic
  disks}},}\ }\href@noop {} {\bibfield  {journal} {\bibinfo  {journal} {Applied
  Physics Letters}\ }\textbf {\bibinfo {volume} {78}},\ \bibinfo {pages} {3848}
  (\bibinfo {year} {2001})}\BibitemShut {NoStop}%
\bibitem [{\citenamefont {Baibich}\ \emph {et~al.}(1988)\citenamefont
  {Baibich}, \citenamefont {Broto}, \citenamefont {Fert}, \citenamefont {{Van
  Dau}}, \citenamefont {Petroff}, \citenamefont {Eitenne}, \citenamefont
  {Creuzet}, \citenamefont {Friederich},\ and\ \citenamefont
  {Chazelas}}]{Baibich1988}%
  \BibitemOpen
  \bibfield  {author} {\bibinfo {author} {\bibfnamefont {M.~N.}\ \bibnamefont
  {Baibich}}, \bibinfo {author} {\bibfnamefont {J.~M.}\ \bibnamefont {Broto}},
  \bibinfo {author} {\bibfnamefont {A.}~\bibnamefont {Fert}}, \bibinfo {author}
  {\bibfnamefont {F.~N.}\ \bibnamefont {{Van Dau}}}, \bibinfo {author}
  {\bibfnamefont {F.}~\bibnamefont {Petroff}}, \bibinfo {author} {\bibfnamefont
  {P.}~\bibnamefont {Eitenne}}, \bibinfo {author} {\bibfnamefont
  {G.}~\bibnamefont {Creuzet}}, \bibinfo {author} {\bibfnamefont
  {A.}~\bibnamefont {Friederich}}, \ and\ \bibinfo {author} {\bibfnamefont
  {J.}~\bibnamefont {Chazelas}},\ }\bibfield  {title} {\enquote {\bibinfo
  {title} {{Giant magnetoresistance of (001)Fe/(001)Cr magnetic
  superlattices}},}\ }\href {\doibase 10.1103/PhysRevLett.61.2472} {\bibfield
  {journal} {\bibinfo  {journal} {Physical Review Letters}\ }\textbf {\bibinfo
  {volume} {61}},\ \bibinfo {pages} {2472} (\bibinfo {year}
  {1988})}\BibitemShut {NoStop}%
\bibitem [{\citenamefont {Binasch}\ \emph {et~al.}(1989)\citenamefont
  {Binasch}, \citenamefont {Gr{\"{u}}nberg}, \citenamefont {Saurenbach},\ and\
  \citenamefont {Zinn}}]{Binasch1989}%
  \BibitemOpen
  \bibfield  {author} {\bibinfo {author} {\bibfnamefont {G.}~\bibnamefont
  {Binasch}}, \bibinfo {author} {\bibfnamefont {P.}~\bibnamefont
  {Gr{\"{u}}nberg}}, \bibinfo {author} {\bibfnamefont {F.}~\bibnamefont
  {Saurenbach}}, \ and\ \bibinfo {author} {\bibfnamefont {W.}~\bibnamefont
  {Zinn}},\ }\bibfield  {title} {\enquote {\bibinfo {title} {{Enhanced
  magnetoresistance in layered magnetic structures with antiferromagnetic
  interlayer exchange}},}\ }\href@noop {} {\bibfield  {journal} {\bibinfo
  {journal} {Physical Review B}\ }\textbf {\bibinfo {volume} {39}},\ \bibinfo
  {pages} {4828} (\bibinfo {year} {1989})}\BibitemShut {NoStop}%
\bibitem [{\citenamefont {Kuepferling}\ \emph {et~al.}(2015)\citenamefont
  {Kuepferling}, \citenamefont {Zullino}, \citenamefont {Sola}, \citenamefont
  {{Van De Wiele}}, \citenamefont {Durin}, \citenamefont {Pasquale},
  \citenamefont {Rott}, \citenamefont {Reiss},\ and\ \citenamefont
  {Bertotti}}]{Kuepferling2015}%
  \BibitemOpen
  \bibfield  {author} {\bibinfo {author} {\bibfnamefont {M.}~\bibnamefont
  {Kuepferling}}, \bibinfo {author} {\bibfnamefont {S.}~\bibnamefont
  {Zullino}}, \bibinfo {author} {\bibfnamefont {A.}~\bibnamefont {Sola}},
  \bibinfo {author} {\bibfnamefont {B.}~\bibnamefont {{Van De Wiele}}},
  \bibinfo {author} {\bibfnamefont {G.}~\bibnamefont {Durin}}, \bibinfo
  {author} {\bibfnamefont {M.}~\bibnamefont {Pasquale}}, \bibinfo {author}
  {\bibfnamefont {K.}~\bibnamefont {Rott}}, \bibinfo {author} {\bibfnamefont
  {G.}~\bibnamefont {Reiss}}, \ and\ \bibinfo {author} {\bibfnamefont
  {G.}~\bibnamefont {Bertotti}},\ }\bibfield  {title} {\enquote {\bibinfo
  {title} {{Vortex dynamics in Co-Fe-B magnetic tunnel junctions in presence of
  defects}},}\ }\href@noop {} {\bibfield  {journal} {\bibinfo  {journal}
  {Journal of Applied Physics}\ }\textbf {\bibinfo {volume} {117}},\ \bibinfo
  {pages} {17E107} (\bibinfo {year} {2015})}\BibitemShut {NoStop}%
\bibitem [{\citenamefont {Julliere}(1975)}]{Julliere1975}%
  \BibitemOpen
  \bibfield  {author} {\bibinfo {author} {\bibfnamefont {M.}~\bibnamefont
  {Julliere}},\ }\bibfield  {title} {\enquote {\bibinfo {title} {{Tunneling
  between ferromagnetic films}},}\ }\href {\doibase
  10.1016/0375-9601(75)90174-7} {\bibfield  {journal} {\bibinfo  {journal}
  {Physics Letters A}\ }\textbf {\bibinfo {volume} {54}},\ \bibinfo {pages}
  {225} (\bibinfo {year} {1975})}\BibitemShut {NoStop}%
\bibitem [{\citenamefont {Scola}\ \emph {et~al.}(2007)\citenamefont {Scola},
  \citenamefont {Polovy}, \citenamefont {Fermon}, \citenamefont
  {Pannetier-Lecœur}, \citenamefont {Feng}, \citenamefont {Fahy},\ and\
  \citenamefont {Coey}}]{Scola2007}%
  \BibitemOpen
  \bibfield  {author} {\bibinfo {author} {\bibfnamefont {J.}~\bibnamefont
  {Scola}}, \bibinfo {author} {\bibfnamefont {H.}~\bibnamefont {Polovy}},
  \bibinfo {author} {\bibfnamefont {C.}~\bibnamefont {Fermon}}, \bibinfo
  {author} {\bibfnamefont {M.}~\bibnamefont {Pannetier-Lecœur}}, \bibinfo
  {author} {\bibfnamefont {G.}~\bibnamefont {Feng}}, \bibinfo {author}
  {\bibfnamefont {K.}~\bibnamefont {Fahy}}, \ and\ \bibinfo {author}
  {\bibfnamefont {J.~M.~D.}\ \bibnamefont {Coey}},\ }\bibfield  {title}
  {\enquote {\bibinfo {title} {Noise in {MgO} barrier magnetic tunnel junctions
  with cofeb electrodes: Influence of annealing temperature},}\ }\href@noop {}
  {\bibfield  {journal} {\bibinfo  {journal} {Applied Physics Letters}\
  }\textbf {\bibinfo {volume} {90}},\ \bibinfo {pages} {252501} (\bibinfo
  {year} {2007})}\BibitemShut {NoStop}%
\end{thebibliography}%

\end{document}